\documentclass[a4paper]{article}
\usepackage[english]{babel}
\usepackage[utf8x]{inputenc}
\usepackage[T1]{fontenc}
\usepackage{amsmath,amsthm,amssymb}
\usepackage{graphicx}
\usepackage[a4paper,top=3cm,bottom=2cm,left=3cm,right=3cm,marginparwidth=1.75cm]{geometry}
\usepackage[font=small,labelfont=bf]{caption}
\usepackage{authblk}
\usepackage[colorlinks=true, allcolors=blue]{hyperref}

\renewcommand\Affilfont{\itshape\small}
\makeatletter
\renewcommand\AB@affilsepx{, \protect\Affilfont}
\makeatother

\title{Putin's peaks: Russian election data revisited}
\author[1]{Dmitry Kobak}
\author[2]{Sergey Shpilkin}
\author[3]{Maxim S. Pshenichnikov}
\affil[1]{University of T\"ubingen, Germany}
\affil[2]{Moscow, Russia}
\affil[3]{Groningen, the Netherlands}

\begin{document}
\maketitle

On Sunday, 18 March 2018, Russian president Vladimir Putin was re-elected to a second consecutive term of six years. It is his fourth term overall, having had a four-year break as Prime Minister in 2008--2012.

The outcome of this year’s election was in little doubt: Putin was the favourite to win, according to all polls. But as CNN was reporting a few weeks before the vote, voter turnout was ``one thing for [Putin] to worry about... It could be embarrassingly low, some polls suggest, and could raise questions about the legitimacy of Putin’s long-running authority'' (\href{http://cnn.it/2H1fAtu}{\nolinkurl{cnn.it/2H1fAtu}}). According to CNN: ``The Kremlin wants the golden 70-70 --- a win of 70\% of the vote with a turnout of 70\% --- to give it a clear mandate and provide it with a riposte to Western leaders who criticize Russia as an autocracy.''

In the event, the official turnout was a seemingly respectable 67.5\%, with Putin securing 77\% of the vote. However, there have been criticisms of the election process --- and doubts have been cast over the validity of the outcome. For example, Golos, an election monitoring organisation, has documented incidents of ballot stuffing at various polling stations, and multiple other violations both before and during the election (\href{http://bit.ly/2HawRD3}{\nolinkurl{bit.ly/2HawRD3}}).  

Criticisms of this sort are not new. Presidential and parliamentary elections in Russia have been accused of being fraudulent since at least the middle of the 2000s. Such accusations have always been denied by Russian officials, but what if there were evidence of this fraud --- fingerprints, as it were --- in the official election data? 

\section*{What do the data suggest?}

In Russia, data from each polling station are freely available online after each election and include the number of registered voters, the number of people who participated in the election, and the number of ballots cast for each candidate. We can apply statistical analysis to these data to see if there are irregularities, which may serve as evidence of falsifications.

Arguably the two most important numbers that describe an election outcome are \textit{turnout percentage} and \textit{leader’s result percentage} (with \textit{leader’s result} referring to Putin during presidential elections and the ruling United Russia party during parliamentary elections). These percentages are not reported in the data sets from individual polling stations but can be calculated from the information provided officially.

We (and others) have previously argued that due to human attraction to round numbers, large-scale attempts to manipulate reported turnout or leader’s results would likely show up as frequent whole (integer) percentages in the election data \cite{kobak2016integer,rozenas2017detecting}. In a previous \textit{Significance} article, we gave the hypothetical example of a polling station with 1755 registered voters \cite{kobak2016statistical}. Here election officials decide to forge the results and report a turnout of 85\%. They choose 85\% because it is a round number which is more appealing than, say, 83.27\%. As we explained: ``To achieve a falsified turnout of 85\%, this polling station needs to report $1755 \times 0.85 = 1492$ ballots cast... Note that the number 1492 is not remarkable in itself; it is only the resulting percentage value (i.e. the 1492/1755 ratio) that is round.'' Other polling stations making similar attempts at fraud may also choose 85\% as their target value, so that when we look at the turnout percentages for all polling stations, we see a noticeable spike in the number of stations with turnout of 85\%. 

In our previous article, we found these integer peaks for elections from 2004 to 2012. Since then, two new elections have been held in Russia: the 2016 parliamentary elections and the 2018 presidential election. These give us the opportunity to again test our hypothesis (see \href{http://github.com/dkobak/elections}{\nolinkurl{github.com/dkobak/elections}} for data and code).

\begin{figure}
\includegraphics[width=\textwidth]{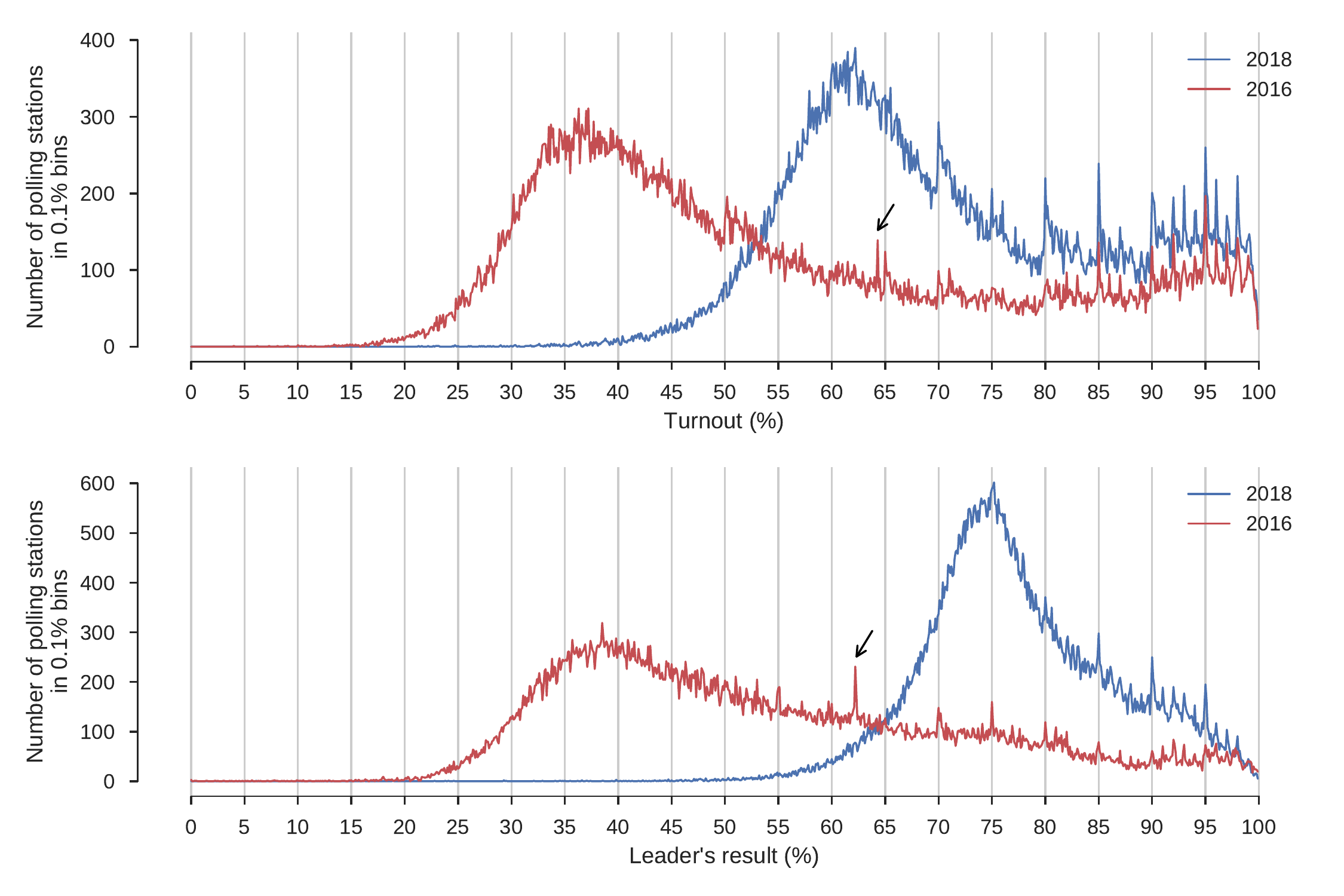}
\caption{Turnout and leader’s result histograms for the 2016 and 2018 Russian elections. Histograms were calculated with 0.1\% bins (centred in such a way that the value at, for example, 50\% corresponds to $50\pm 0.05\%$). To avoid any artefacts due to integer division, we added a random number sampled from the uniform distribution $\mathcal U(-0.5, 0.5)$ to the numerator of each turnout or leader’s result fraction. Polling stations with 100\% turnout were excluded from this figure. Black arrows mark peaks attributable to the city of Saratov; see ``Regional peaks''.}
\label{fig:peaks}
\end{figure}

Figure~\ref{fig:peaks} shows histograms for turnout and leader's result at polling stations in the two most recent elections. As with previous elections, sharp periodic peaks are clearly visible at integer values (such as 91\%, 92\% and 93\%) and at round integer values (such as 80\%, 85\% and 90\%), rather than fractional values (such as 91.3\%). 

\section*{Comparing peak prevalence}

Having again identified peaks at integer values, we wondered whether there had been any change in their prevalence over time, particularly given the scrutiny applied to past elections by the media and academics? To address this question, we compared the number of polling stations with integer percentages for turnout or leader’s result against the number that would be expected by chance. The expected values were computed by Monte Carlo simulations of election results using the binomial distribution of ballots at every polling station. 

\begin{figure}
\begin{center}
\includegraphics[width=.6\textwidth]{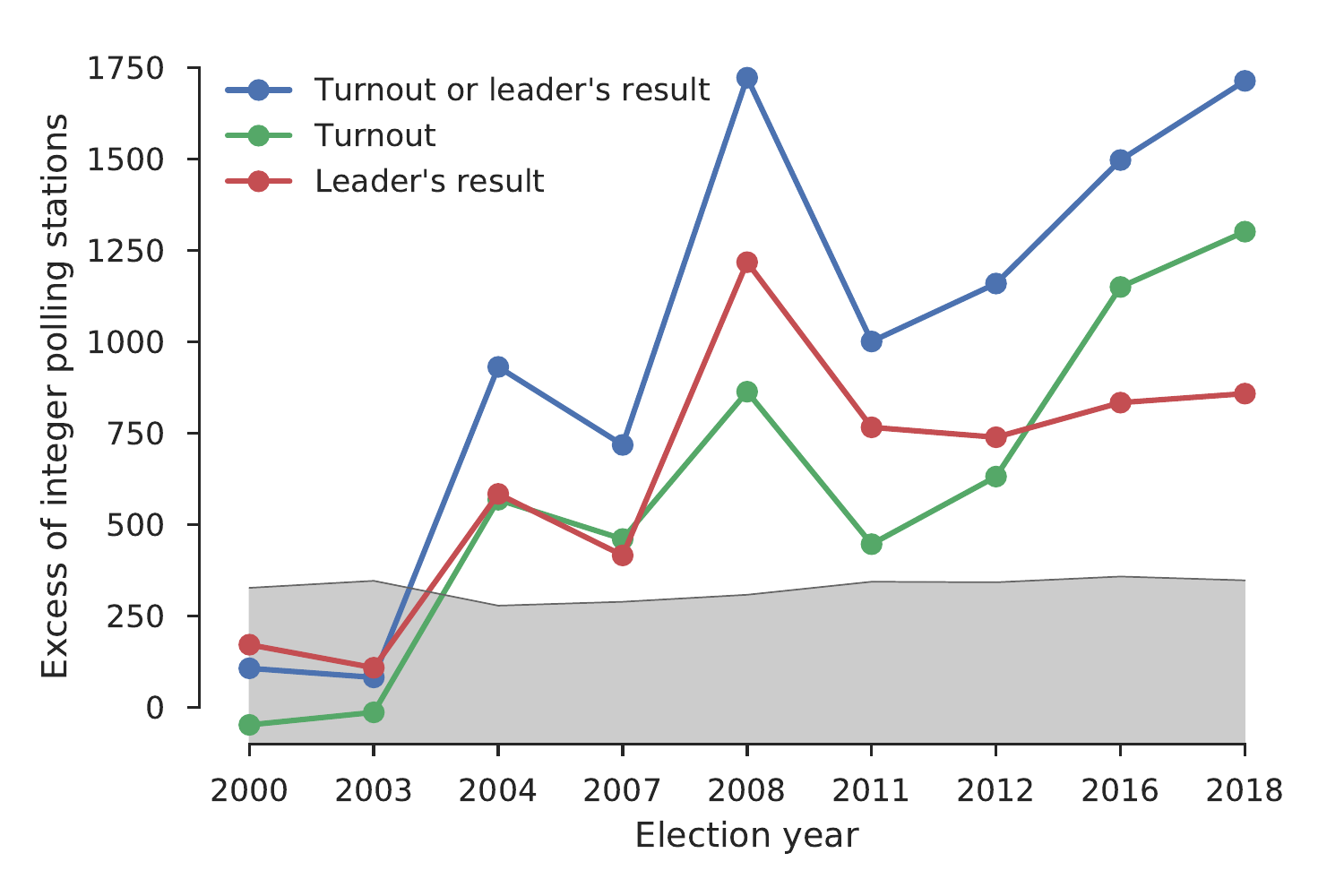}
\end{center}
\caption{Integer anomalies for all Russian elections from 2000 to 2018. The anomaly is defined as the excess of polling stations with integer turnout (green), leader's result (red) and any of the two (blue), compared to the expected values from binomial Monte Carlo simulations. Grey shading shows the interval within the 99.9 percentile of the Monte Carlo values (for the joint anomaly shown in blue), corresponding to $p<0.001$ for any value above it. See \cite{kobak2016integer} for computational details.}
\label{fig:anomalies}
\end{figure}

Figure~\ref{fig:anomalies} shows the excess of polling stations with integer values for all elections between 2000 and 2018. Before 2004, the number of polling stations with integer values for either turnout or leader’s result (the blue curve) was close to that expected by chance, indicated by the null hypothesis value of 0 excess on the $y$-axis. From 2004, the excess of integer polling stations increased, spiking in 2008 and dropping back in 2011. Since then, it has steadily increased to levels last seen in 2008. 

The excess of polling stations with integer values for turnout or leader’s result at both the 2016 and 2018 elections is far beyond what we would expect to find as the result of random chance. The grey shading in the figure shows intervals until the 99.9 percentile of values obtained in our Monte Carlo simulations; any individual value above this shaded area has a probability of less than one in a thousand ($p<0.001$) under the statistical model used. 

It is also instructive to look separately at the excess of polling stations with integer values for turnout (green curve) and leader's result (red curve). For the latter, the excess has stayed constant since 2011. However, the former has kept growing and reached its historic maximum in 2018. If these results are taken as evidence of electoral data manipulation, this might suggest that recent efforts to manipulate results have focused on turnout, presumably because the leader's result was believed to be high enough already.

\section*{Regional peaks}

As we have shown previously \cite{kobak2016integer}, integer peaks in the election data do not originate uniformly across all parts of the country; they are mostly localised in the same administrative regions, providing additional evidence in support of our hypothesis that these are not natural phenomena. Specific integer peaks can sometimes be traced to a particular city, or even an electoral constituency within a city, where turnout and/or leader’s result are nearly identical at a large number of polling stations \cite{kobak2016statistical}.

The most prominent example from the last two elections was the city of Saratov in 2016. Its polling stations are the sole contributor to the sharp turnout peak at 64.3\% and the leader's result peak at 62.2\%, both visible in Figure~\ref{fig:peaks}. These peaks are not integer and so are not counted towards the anomalies computed for Figure~\ref{fig:anomalies}. Curiously, their product --- showing the fraction of leader’s votes with respect to the total number of registered voters --- is $0.643 \times 0.622 = 0.400$, which is exactly round.

\newpage
\section*{Acknowledgements}
We thank Brian Tarran (\textit{Significance} magazine) for his valuable help in editing this article.

\bibliographystyle{alpha}
\bibliography{main}

\begin{thebibliography}{KSP16b}

\bibitem[KSP16a]{kobak2016integer}
Dmitry Kobak, Sergey Shpilkin, and Maxim~S Pshenichnikov.
\newblock Integer percentages as electoral falsification fingerprints.
\newblock {\em The Annals of Applied Statistics}, 10(1):54--73, 2016.

\bibitem[KSP16b]{kobak2016statistical}
Dmitry Kobak, Sergey Shpilkin, and Maxim~S Pshenichnikov.
\newblock Statistical fingerprints of electoral fraud?
\newblock {\em Significance}, 13(4):20--23, 2016.

\bibitem[Roz17]{rozenas2017detecting}
Arturas Rozenas.
\newblock Detecting election fraud from irregularities in vote-share
  distributions.
\newblock {\em Political Analysis}, 25(1):41--56, 2017.

\end{thebibliography}

\end{document}